\documentclass[aps,pre,twocolumn,superscriptaddress,nofootinbib]{revtex4-2}

\usepackage{amsmath,amssymb,bm,mathtools}
\usepackage{physics}
\usepackage{microtype}
\usepackage{hyperref}
\usepackage{booktabs}

\newcommand{\uu}{\bm{u}}

\newcommand{\II}{\bm{I}}

\newcommand{\bb}{\bm{b}}
\newcommand{\Th}{\bm{\Theta}}
\newcommand{\Op}{\operatorname{Op}}
\newcommand{\Rey}{\mathrm{Re}}
\newcommand{\trc}{\operatorname{tr}}

\begin{document}

\title{A Tensorial Extension of Fractional Dissipation at Inertial Scales}

\author{Jos\'e I.H. L\'opez}
\affiliation{Department of Mechanical Engineering, University of S\~ao Paulo (USP), Brazil, Av. Prof. Mello Moraes, 2231, S\~ao Paulo, 05508-030, SP, Brazil}
\email{jihlpez@gmail.com}

\date{September 26, 2026}

\begin{abstract}
We formulate a scale-dependent tensorial extension of a fractional dissipation operator motivated by a macroscopic Navier--Stokes correspondence. The starting point is a fractional model in which the order of dissipation decreases continuously from the Laplacian value toward the Kolmogorov value as a local Reynolds number increases. At the critical value of the local Reynolds number, the model reduces to an isotropic operator proportional to $(-\Delta)^{1/3}$. We then ask whether the same macroscopic description can be continued into a regime in which velocity fields have Onsager-critical or subcritical regularity without introducing an unrelated closure or an additional isotropic dissipation scale.

In addition, we extend the original AFNS free-energy construction by one nonnegative concentration order parameter $c$. The new term is quadratic in $c$ and linearly coupled to the defect beta function, so that minimization leaves the original AFNS free energy exactly unchanged when the concentration branch is inactive, while a positive concentration amplitude appears only for $\beta_D<0$. This constrained construction uses the standard Landau free-energy logic and the Kuhn--Tucker conditions for the boundary $c\ge0$ \cite{Landau1937,GinzburgLandau1950,KuhnTucker1951}. The resulting three-regime mosaic is regular/NS, AFNS/inertial, and concentrated/anomalous, with the onset of concentration at $h=1/3$.

The construction uses the coarse-grained stress $\tau_\ell=\overline{\uu\otimes\uu}_\ell-\bar{\uu}_\ell\otimes\bar{\uu}_\ell$ and its critical normalization $\Th_\ell=\ell^{-2/3}\tau_\ell$. The normalized anisotropy tensor $\bb_\ell=\Th_\ell/\trc\Th_\ell-\II/3$ provides the lowest-order objective tensorial information available at the critical scale, while the scale derivative of the Duchon--Robert energy defect, $\beta_D=d\ln|D_\ell|/d\ln\ell=3(h-1/3)$, supplies a scalar measure of departure from the Onsager threshold. We propose the minimal constitutive deformation $\exp(\beta_D\bb_\ell)$ of the critical operator: in Weyl quantization, the isotropic fractional symbol $\eta_c|\xi|^{2/3}$ is dressed by this exponential tensorial factor, giving a pseudodifferential operator that reduces to the scalar critical operator whenever $\beta_D=0$. The exponential parametrization guarantees positive eigenvalues and unit determinant for symmetric traceless $\bb_\ell$, so the deformation redistributes dissipation among directions without introducing a new isotropic amplitude, and the critical limit $\beta_D\to0$ exactly recovers $\eta_c(-\Delta)^{1/3}$. We emphasize that the tensorial structure is constrained by established representation and realizability results, whereas the exponential closure is a hypothesis, not a derivation from microscopic fluid mechanics. We therefore formulate direct falsifiability tests based on DNS and experiments rather than claiming that the proposed operator is established turbulence physics.
\end{abstract}

\maketitle

\section{Introduction}

The macroscopic description of turbulence is unusual among continuum theories because several descriptions remain simultaneously useful while emphasizing different levels of the same multiscale dynamics. The incompressible Navier--Stokes equations provide the standard local balance laws, while Kolmogorov's theory identifies a distinguished inertial-range scaling and the $4/5$ law supplies an exact third-order constraint in homogeneous, isotropic turbulence. Onsager's analysis and the subsequent theory of weak Euler solutions further show that the regularity threshold $h=1/3$ is not merely a phenomenological exponent: it separates velocity fields for which the nonlinear energy flux vanishes from fields for which a nonzero inertial defect can persist \cite{Onsager1949,Kolmogorov1941a,Kolmogorov1941b,DuchonRobert2000}. These facts motivate the question addressed here: can a macroscopic dissipative operator be continued through this threshold without replacing the original dynamics by an unrelated phenomenological closure?

Our starting point is a fractional extension of the incompressible Navier--Stokes equation,
\begin{equation}
\partial_t\uu+(\uu\cdot\nabla)\uu+\nabla p+\eta(s,\Rey_\ell)(-\Delta)^s\uu=0,
\qquad \nabla\cdot\uu=0,
\label{eq:afns}
\end{equation}
where the order $s$ is not fixed but is a function of a local Reynolds number. This adaptive fractional Navier--Stokes (AFNS) framework was introduced in Ref.~\cite{LopezAFNS2026}. The specific interpolation considered below is
\begin{equation}
 s(\Rey_\ell)=\frac13+\delta(\Rey_\ell),
\qquad
\delta(\Rey_\ell)=\frac{2/3}{1+(\Rey_\ell/\Rey_c)^\gamma},
\label{eq:sdelta}
\end{equation}
with $\gamma>0$. The form in Eq.~\eqref{eq:sdelta} is used here as the macroscopic AFNS realization; the present paper does not claim that its particular interpolation is uniquely derived from microscopic fluid dynamics.

The essential feature is the correspondence trajectory. At the beginning of the Reynolds-number evolution, $s$ approaches the Navier--Stokes value $1$. At large local Reynolds number, the order approaches $1/3$. The distinguished scale $\Rey_c$ therefore separates an ordinary Laplacian regime from a lower-order fractional regime. The coefficient is chosen so that the critical operator has a finite amplitude,
\begin{equation}
\eta\longrightarrow \frac{1}{\Rey_c},
\qquad
s\longrightarrow\frac13,
\label{eq:criticalamplitude}
\end{equation}
so that the limiting isotropic amplitude is \begin{equation*}\eta_c\equiv\eta(1/3,\infty)=\frac{1}{\Rey_c},\end{equation*} and the limiting isotropic operator is $\eta_c(-\Delta)^{1/3}$.

The new question is not whether fractional dissipation can be written down. It is whether the same description can contain the anomalous side of the Onsager threshold. A purely scalar fractional operator cannot encode directional information generated by the flow itself. Conversely, importing a standard Reynolds-stress or large-eddy closure would change the purpose of the construction: it would introduce a model designed to represent unresolved stress rather than a constitutive continuation of the AFNS operator.

We therefore proceed in the opposite direction. Instead of postulating an effective viscosity tensor, we ask what tensorial information is already present in the coarse-grained velocity field at the critical scale. This leads to the subscale stress
\begin{equation}
\tau_\ell=\overline{\uu\otimes\uu}_\ell-\bar{\uu}_\ell\otimes\bar{\uu}_\ell,
\label{eq:tau}
\end{equation}
and to its critical normalization
\begin{equation}
\Th_\ell=\ell^{-2/3}\tau_\ell.
\label{eq:theta}
\end{equation}
If $|\delta_\ell u|\sim \ell^h$, then $\tau_\ell\sim\ell^{2h}$ and hence $\Th_\ell\sim\ell^{2h-2/3}$. The normalization is therefore neutral at $h=1/3$: above the threshold it decays, at the threshold it is of order unity, and below the threshold it grows. This is precisely the scale at which a tensorial constitutive deformation can become relevant without being introduced by hand.

The corresponding scalar transfer information is supplied by the inertial energy defect. The Duchon--Robert construction identifies a local defect distribution associated with the failure of smooth solutions to satisfy the classical energy balance \cite{DuchonRobert2000}. Under the scaling $\delta_\ell u\sim\ell^h$, the defect scales as
\begin{equation}
D_\ell\sim\ell^{3h-1}.
\label{eq:Dscaling}
\end{equation}
This gives the defect beta function
\begin{equation}
\beta_D\equiv\frac{d\ln|D_\ell|}{d\ln\ell}=3h-1.
\label{eq:betad}
\end{equation}
The threshold $h=1/3$ is therefore a fixed point of the defect scaling, not necessarily a fixed point of the AFNS order itself.

A second scale flow is present. If
\begin{equation}
\kappa\equiv\frac{d\ln\Rey_\ell}{d\ln\ell},
\label{eq:kappa}
\end{equation}
then the AFNS relation used here is
\begin{equation}
h-\frac13=\kappa-2\delta,
\label{eq:hdeltakappa}
\end{equation}
so that
\begin{equation}
\beta_D=3(\kappa-2\delta).
\label{eq:betadelta}
\end{equation}
This relation is important because it prevents a common conceptual conflation: $\beta_D$ is the beta function of the defect scaling, whereas
\begin{equation}
\beta_\delta=\frac{d\delta}{d\ln\ell}
\label{eq:betadelta2}
\end{equation}
is the beta function of the fractional-order coupling. They are distinct flows.

The $4/5$ law provides a particularly sharp statistical anchor. In homogeneous and isotropic stationary turbulence, the mean longitudinal third-order structure function satisfies $\langle(\delta u_L)^3\rangle=-4\varepsilon\ell/5$ in the inertial range, and DNS studies use this relation as a stringent benchmark \cite{TaylorKurienEyink2003}. Rigorous results also establish versions of the law for forced three-dimensional Navier--Stokes equations under suitable assumptions in the vanishing-viscosity setting \cite{HofmanovaEtAl2023}. The law constrains the mean flux; it does not imply that a local defect $D_\ell(x)$ is pointwise constant. This distinction is essential for an intermittency-sensitive construction.

The tensorial part of the problem has a long classical history. Objectivity and isotropy constrain constitutive tensor functions through finite representation theorems. In turbulence modeling, Pope demonstrated that an effective-viscosity relation depending on the velocity-gradient tensor can be expressed as a finite tensor polynomial with scalar coefficient functions of invariants \cite{Pope1975,Pope2000}. Reynolds-stress modeling further provides realizability constraints on the anisotropy tensor, traditionally represented by the Lumley invariant domain and, in a geometrically transparent form, by the barycentric map \cite{Lumley1978,BanerjeeEtAl2007}. These results do not determine the physics of the present problem, but they sharply restrict its admissible tensorial structure.

Our objective is therefore deliberately modest. We construct the minimal tensorial continuation compatible with (i) the AFNS critical operator, (ii) rotational covariance, (iii) symmetry and tracelessness of the anisotropic generator, (iv) positivity of the principal dissipative symbol, and (v) scale consistency. The result is a proposal, not a completed theory of turbulence. Its value, if any, lies in producing a compact operator with explicit experimental and numerical failure modes.

The question of concentration has a long history in fluid mechanics. Leray's self-similar analysis already identified concentration as a natural route to possible singular behavior in Navier--Stokes flow, while Scheffer's work on partial regularity and Hausdorff dimension, followed by the Caffarelli--Kohn--Nirenberg theory, placed quantitative geometric restrictions on possible singular sets \cite{Leray1934,Scheffer1976,Scheffer1977,CKN1982}. More recent constructions of imploding or concentrating solutions in related fluid equations have made the geometry of concentration increasingly explicit \cite{MerleEtAl2022,BuckmasterEtAl2023,CaoLaboraEtAl2025}. This literature motivates an inverse question: if a mathematical construction can concentrate dynamically on increasingly small structures, what macroscopic constitutive signature would distinguish such concentration from the behavior accessible to physical turbulent fluids? In September 2026, contemporaneous finite-time blow-up constructions by Alp\"oge and Buckmaster and a separate Navier--Stokes construction released by OpenAI made this question especially concrete \cite{AlpogeBuckmaster2026,OpenAI2026}. We cite these works only to acknowledge the motivation supplied by the recent concentration literature; no specific theorem, ansatz, parameter value, or construction from either work is used in the derivation below. Our hypothesis is instead the inverse one: physical fluids may suppress or redistribute concentration through a scale-dependent tensorial response. Whether that hypothesis is correct is an empirical question.

\section{From AFNS to the critical scale}

Equation~\eqref{eq:afns} may be viewed as a trajectory in operator space. The fractional order is
\begin{equation}
1\ge s(\Rey_\ell)>\frac13,
\end{equation}
with
\begin{equation}
\delta=s-\frac13.
\end{equation}
Under Kolmogorov scaling, $\Rey_\ell\propto\ell^{4/3}$, so Eq.~\eqref{eq:sdelta} gives
\begin{equation}
\beta_\delta
=\frac{d\delta}{d\ln\ell}
=-\frac{4\gamma}{3}\delta\left(1-\frac32\delta\right).
\label{eq:betadeltaK41}
\end{equation}
The two zeros correspond formally to $s=1$ and $s=1/3$. We do not identify them here with specific field-theoretic universality classes; doing so would require a fully defined renormalization transformation on the theory space.

At the critical end of the AFNS trajectory,
\begin{equation}
\delta\rightarrow0,
\qquad s\rightarrow\frac13,
\qquad \eta\rightarrow\frac1{\Rey_c},
\end{equation}
so the operator becomes
\begin{equation}
\mathcal L_c=\eta_c(-\Delta)^{1/3}.
\label{eq:Lc}
\end{equation}
The question is what additional information is needed to continue this operator into $h\le1/3$.

\section{The critical tensor supplied by coarse graining}

Let $G_\ell$ be a normalized smooth filter and denote filtered quantities by overbars. The exact coarse-grained nonlinear stress is Eq.~\eqref{eq:tau}. It is symmetric because $u_i u_j=u_j u_i$. It is also positive semidefinite in the usual pointwise sense when written as the filtered covariance of velocity components. The tensor does not assume a particular flow geometry.

The critical normalization Eq.~\eqref{eq:theta} removes the Kolmogorov dimensional scaling. If $\delta_\ell u\sim\ell^h$, then
\begin{equation}
\Theta_\ell\sim\ell^{2h-2/3}.
\end{equation}
Thus
\begin{equation}
\begin{cases}
\Theta_\ell\to0,&h>1/3,\\
\Theta_\ell=O(1),&h=1/3,\\
\|\Theta_\ell\|\to\infty,&h<1/3,
\end{cases}
\label{eq:theta-regimes}
\end{equation}
where the last line is a scaling statement and does not assert existence of a pointwise limit.

The normalized shape tensor is
\begin{equation}
Q_\ell=\frac{\Theta_\ell}{\trc\Theta_\ell},
\qquad
\bb_\ell=Q_\ell-\frac13\II.
\label{eq:bdef}
\end{equation}
When $\trc\Theta_\ell>0$, $Q_\ell$ is symmetric, positive semidefinite, and has unit trace, while $\bb_\ell$ is symmetric and traceless. Its admissible eigenvalue domain is constrained by Reynolds-stress realizability. The Lumley invariant domain and its barycentric representation provide established ways of visualizing this constraint \cite{Lumley1978,BanerjeeEtAl2007}.

The use of $\bb_\ell$ rather than the magnitude of $\Theta_\ell$ is deliberate. The AFNS amplitude already fixes the isotropic scale at the critical point. The role of $\bb_\ell$ is to encode shape, orientation, and anisotropic redistribution, not to introduce a second scalar dissipation amplitude.

\section{The scalar distance from the Onsager threshold}

The inertial energy flux is cubic in velocity increments. A representative coarse-grained flux has the scaling
\begin{equation}
\Pi_\ell\sim \frac{|\delta_\ell u|^3}{\ell}\sim\ell^{3h-1}.
\end{equation}
The same exponent governs the Duchon--Robert defect. Hence Eq.~\eqref{eq:betad} follows.

Combining it with Eq.~\eqref{eq:hdeltakappa} gives
\begin{equation}
\boxed{\beta_D=3(\kappa-2\delta)}.
\end{equation}
This quantity has a simple interpretation. At $\beta_D=0$, the defect is scale-invariant. At $\beta_D>0$, the defect decreases toward smaller scales under the corresponding power-law description; at $\beta_D<0$, the defect is amplified toward smaller scales. The interpretation is local and scaling-based; the exact $4/5$ law constrains suitable averages rather than every realization.

The key point is that the AFNS order flow and the defect flow are different. The operator coupling follows $\beta_\delta$, whereas the anomalous sector is detected by $\beta_D$. The critical manifold is therefore naturally described by
\begin{equation}
\beta_D=0\quad\Longleftrightarrow\quad h=\frac13\quad\Longleftrightarrow\quad\kappa=2\delta.
\label{eq:criticalmanifold}
\end{equation}

\section{Tensorial constitutive structure}

The desired correction must be an objective tensor function of dimensionless local information. If one uses the full velocity-gradient tensor, classical representation theory implies a finite tensor basis constructed from the strain-rate and rotation tensors and scalar invariants. Pope's formulation is an established example of this reduction \cite{Pope1975}.

For the present minimal construction, however, the normalized coarse-grained stress already supplies a symmetric traceless tensor $\bb$. By the Cayley--Hamilton theorem in three dimensions, polynomial functions of one $3\times3$ tensor reduce to the basis generated by $\II$, $\bb$, and $\bb^2$. Removing the isotropic component leaves
\begin{equation}
\bb,
\qquad
\bb^2-\frac13\trc(\bb^2)\II.
\label{eq:minbasis}
\end{equation}
Therefore the most general polynomial traceless isotropic function of $\bb$ has the form
\begin{equation}
\mathsf B
=G_1(\delta,\beta_D,II_b,III_b)\bb
+G_2(\delta,\beta_D,II_b,III_b)
\left[\bb^2-\frac13II_b\II\right],
\label{eq:Bgeneral}
\end{equation}
where $II_b=\trc(\bb^2)$ and $III_b=\trc(\bb^3)$.

Equation~\eqref{eq:Bgeneral} is a representation result, not a turbulence closure. It reduces the admissible tensorial freedom but does not select the scalar functions. We now make the additional minimality assumption that the first departure from the critical point is linear in the anisotropy and contains no independent scalar constitutive function. This selects
\begin{equation}
\boxed{\mathsf B_\ell=\bb_\ell.}
\label{eq:Bminimal}
\end{equation}
This is the principal modeling hypothesis of the paper. It is not derived from Navier--Stokes dynamics. It is chosen because it is the unique lowest-order generator directly available from the critical coarse-grained stress once isotropic amplitude and tracelessness have been separated.

\section{A positive tensorial deformation}

A linear correction $\II+\beta_D\bb$ is sufficient close to the critical point, but it does not guarantee positivity for arbitrarily large $|\beta_D|$. We therefore use the exponential parametrization
\begin{equation}
\mathsf M_\ell=\exp(\beta_D\bb_\ell).
\label{eq:M}
\end{equation}
Because $\bb_\ell$ is symmetric, $\mathsf M_\ell$ is symmetric positive definite. Because $\trc\bb_\ell=0$,
\begin{equation}
\det\mathsf M_\ell
=\exp(\beta_D\trc\bb_\ell)=1.
\label{eq:detM}
\end{equation}
Thus the deformation redistributes the principal action of the operator without introducing an additional isotropic volume factor.

The small-defect expansion is
\begin{equation}
\mathsf M_\ell
=\II+\beta_D\bb_\ell+O(\beta_D^2).
\label{eq:smallM}
\end{equation}
Since $\beta_D=3(h-1/3)$, the first correction is directly proportional to the distance from the Onsager threshold.

\section{Variational extension and the generalized AFNS mosaic}

The original AFNS free-energy construction contains two competing contributions: a logarithmic cost for maintaining regularity and an entropic mixing term for the regularity fraction $f$ \cite{LopezAFNS2026}. We now ask for the smallest extension capable of representing the concentrated side of the Onsager threshold without modifying the original functional when concentration is absent. The appropriate new variable is therefore not a third statistical probability, but a nonnegative dimensionless concentration order parameter $c\ge0$. This distinction is important: treating $c$ as a third probability would introduce an additional mixing entropy and would generically prevent a sharp $c=0$ branch for finite control parameters.

We define the generalized local free energy by
\begin{equation}
\boxed{\begin{aligned}
\mathcal F(f,c;\rho,\kappa)={}&
\epsilon_0\ln\rho\,f
+k_BT_{\rm top}\left[f\ln f+(1-f)\ln(1-f)\right]\\
&+\frac{a}{2}c^2+g\left(3\kappa-4f\right)c,
\end{aligned}}
\label{eq:freegeneral}
\end{equation}
with
\begin{equation}
0\le f\le1,\qquad c\ge0,
\qquad \rho=\frac{\Rey_\ell}{\Rey_c},
\end{equation}
and $a>0$, $g>0$. The variable $c$ is dimensionless; $a$ and $g$ have the same free-energy units as $\epsilon_0$. The coefficient $a$ is a concentration stiffness: it penalizes arbitrarily large concentration amplitudes and makes the variational problem bounded in the $c$ direction. The coefficient $g$ is the coupling between concentration and the defect-scaling drive. No new Reynolds-number scale is introduced by either parameter.

The factor multiplying $c$ is not an additional phenomenological exponent. Using $\delta=2f/3$ and Eq.~\eqref{eq:betad},
\begin{equation}
3\kappa-4f=3(\kappa-2\delta)=\beta_D.
\label{eq:betaf}
\end{equation}
Thus the new term is simply $g\beta_Dc$. The generalized functional can therefore be written equivalently as
\begin{equation}
\mathcal F=F_0(f;\rho)+\frac{a}{2}c^2+g\beta_Dc,
\label{eq:freecompact}
\end{equation}
where $F_0$ is exactly the original AFNS free energy.

Because $c$ is restricted to the half-line, its equilibrium condition is a Kuhn--Tucker condition rather than an unconstrained stationarity equation \cite{KuhnTucker1951}. It gives
\begin{equation}
\boxed{c_* = \frac{g}{a}\,[-\beta_D]_+,}
\label{eq:cstar}
\end{equation}
where $[x]_+=\max(x,0)$ and $[-x]_+=\max(-x,0)$. Consequently,
\begin{equation}
\begin{cases}
c_*=0,&\beta_D\ge0,\\
c_*>0,&\beta_D<0.
\end{cases}
\label{eq:cbranches}
\end{equation}
Since $\beta_D=3h-1$, the onset of the concentration branch is exactly
\begin{equation}
\boxed{\beta_D=0\quad\Longleftrightarrow\quad h=\frac13.}
\label{eq:concentrationthreshold}
\end{equation}
No condition $h=1/3$ has been imposed independently; it follows from the constrained minimization and the previously defined defect beta function.

The regression to the original AFNS theory is exact. Whenever $\beta_D\ge0$, Eq.~\eqref{eq:cstar} gives $c_*=0$, and Eq.~\eqref{eq:freegeneral} reduces identically to the original two-state AFNS functional. Its minimization therefore gives
\begin{equation}
f(\rho)=\frac{1}{1+\rho^\gamma},
\qquad
s(\rho)=\frac13+\frac{2/3}{1+\rho^\gamma},
\label{eq:mobiusregression}
\end{equation}
with no renormalization of $\Rey_c$ or $\gamma$ by the new branch.

Eliminating $c$ gives a useful closed form for the effective free energy,
\begin{equation}
\boxed{
\mathcal F_{\rm eff}(f;\rho,\kappa)
=F_0(f;\rho)-\frac{g^2}{2a}[\beta_D]_-^2,
}
\label{eq:freeeffective}
\end{equation}
where $[x]_-=[-x]_+$. The concentration correction is therefore identically zero on the nonconcentrated branch and lowers the free energy only when $\beta_D<0$. On that branch the $f$-stationarity condition becomes
\begin{equation}
\epsilon_0\ln\rho+k_BT_{\rm top}\ln\frac{f}{1-f}-4gc_*=0.
\label{eq:fconcentrated}
\end{equation}
Thus concentration feeds back on the spectral order only after the threshold has been crossed; it does not alter the original Möbius trajectory beforehand.

The resulting macroscopic mosaic is therefore
\begin{equation}
\boxed{
\begin{array}{ccl}
\text{regular / NS} &\longrightarrow& \text{AFNS / inertial mosaic}\\
 c_*=0,\;s\simeq1 && c_*=0,\;1/3<s<1\\[1mm]
 &\longrightarrow& \text{concentrated / anomalous branch}\\
 && c_*>0,\;h<1/3.
\end{array}}
\label{eq:mosaic}
\end{equation}
The common boundary is selected by $\beta_D=0$. The functional itself does not, however, prove that $s=1/3$ and $h=1/3$ are reached at exactly the same point in $\rho$; that coincidence additionally depends on the asymptotic behavior of $\kappa(\rho)$. We therefore retain it as a testable consequence rather than an imposed identity.

The role of the new parameters is correspondingly limited. The concentration stiffness $a$ controls the amplitude of the newly available branch, while $g$ controls its coupling to the defect flow. Their ratio determines the equilibrium concentration scale, $c_*\propto g/a$. Neither parameter changes the AFNS transition scale or the critical fractional amplitude. This is the intended minimality of the extension.

The concentration variable is not inserted as an additional multiplicative coefficient in the dissipative operator. Doing so would introduce a new isotropic constitutive amplitude and would obscure the purpose of the extension. Instead, $c$ selects the variational branch, while the critical operator continues to be deformed by the already defined pair $(\beta_D,\bb_\ell)$. The generalized AFNS equation therefore retains the same principal-symbol form, but its constitutive state now includes the concentration branch selected by Eq.~\eqref{eq:cstar}.

\section{The proposed operator}

The final symbol is
\begin{equation}
\boxed{
P_{\rm IR}(x,\xi,t)
=\eta_c|\xi|^{2/3}
\exp\!\left[\beta_D(x,t)\bb_\ell(x,t)\right].
}
\label{eq:symbol}
\end{equation}
For variable coefficients, the corresponding operator is taken in Weyl quantization,
\begin{equation}
\boxed{
\mathcal L_{\rm IR}\uu
=\eta_c
\Op^W\!\left[
|\xi|^{2/3}e^{\beta_D\bb_\ell}
\right]\uu.
}
\label{eq:LIR}
\end{equation}
The proposed macroscopic equation is therefore
\begin{equation}
\boxed{\begin{aligned}
\partial_t\uu+(\uu\cdot\nabla)\uu+\nabla p
+\eta_c\Op^W\!\left[|\xi|^{2/3}e^{\beta_D\bb_\ell}\right]\uu&=0,\\
\nabla\cdot\uu&=0.
\end{aligned}}
\label{eq:main}
\end{equation}
with
\begin{equation}
\bb_\ell
=\frac{\ell^{-2/3}\left(\overline{\uu\otimes\uu}_\ell-\bar\uu_\ell\otimes\bar\uu_\ell\right)}
{\trc\left\{\ell^{-2/3}\left(\overline{\uu\otimes\uu}_\ell-\bar\uu_\ell\otimes\bar\uu_\ell\right)\right\}}
-\frac13\II
\label{eq:bexplicit}
\end{equation}
and
\begin{equation}
\beta_D=3\left(\frac{d\ln\Rey_\ell}{d\ln\ell}-2\delta(\Rey_\ell)\right).
\label{eq:betaexplicit}
\end{equation}

The relation to AFNS is transparent. The original dissipative term is
\begin{equation}
\eta(s,\Rey_\ell)(-\Delta)^s\uu,
\end{equation}
whereas the continuation beyond the critical order is
\begin{equation}
\eta_c\Op^W\!\left[|\xi|^{2/3}e^{\beta_D\bb_\ell}\right]\uu.
\end{equation}
The first stage changes the scalar spectral order. The second stage retains the critical order and permits a tensorial deformation driven by the defect flow.

At the critical manifold,
\begin{equation}
\beta_D=0
\quad\Rightarrow\quad
P_{\rm IR}=\eta_c|\xi|^{2/3}\II,
\end{equation}
so that
\begin{equation}
\mathcal L_{\rm IR}=\eta_c(-\Delta)^{1/3}.
\end{equation}
The anomalous regime is therefore not represented by replacing the theory at $h=1/3$; it is represented by a deformation of the same critical operator for $h<1/3$.

\section{Scale consistency and relation to dynamic closures}

The construction has a natural two-scale interpretation. Under a scale change $\ell\mapsto a\ell$, the fields $\delta$, $\beta_D$, and $\bb$ are transformed, while the functional form
\begin{equation}
\mathsf M(\delta,\beta_D,\bb)=e^{\beta_D\bb}
\end{equation}
remains unchanged. This is the sense in which the proposal is RG-covariant: no new coefficient is introduced at the second scale.

This idea is related in spirit to real-space renormalization, where a change of scale induces a flow of effective parameters and fixed points \cite{EfratiEtAl2014,HasenfratzWitzel2019}. We do not claim to have constructed a Wilsonian RG transformation of the Navier--Stokes functional integral. Rather, the notation identifies two scale derivatives within the macroscopic theory and demands constitutive consistency under a change of coarse-graining scale.

The construction is also deliberately distinct from large-eddy-simulation subgrid-scale closures, in which an eddy-viscosity coefficient is either fixed a priori, as in Schumann's original subgrid model \cite{Schumann1975}, or determined dynamically from an algebraic identity involving two filter levels, as in the procedure of Germano \emph{et al.} \cite{GermanoEtAl1991}. That two-scale methodology provides a useful precedent for demanding scale consistency, but Eq.~\eqref{eq:main} is not proposed as an LES eddy-viscosity closure. The tensor $\bb_\ell$ enters as a constitutive deformation of the critical fractional operator, while the scalar $\beta_D$ is fixed by the defect scaling relation.

\section{Energy and realizability constraints}

At the level of the principal symbol, Eq.~\eqref{eq:symbol} is positive definite. If $v\in\mathbb R^3$ and $v\ne0$, then
\begin{equation}
v^T P_{\rm IR}v
=\eta_c|\xi|^{2/3}
 v^T e^{\beta_D\bb}v>0.
\end{equation}
This establishes principal coercivity of the proposed dissipative symbol. For spatially varying coefficients, a complete energy estimate must account for lower-order terms generated by quantization and commutators; the present construction does not claim that those estimates are automatic.

The realizability condition on $\bb$ is independent of this operator coercivity. The Reynolds-stress tensor must remain positive semidefinite, which restricts the invariants of $\bb$ to the classical realizability domain \cite{Lumley1978,BanerjeeEtAl2007}. If a measured or simulated trajectory required values of $\bb$ outside this domain, the proposed identification of the constitutive tensor with a normalized coarse-grained covariance would fail before the operator is even constructed.

Energy conservation at the inertial-transfer level enters through $\beta_D$, not through an arbitrary trace component of $\mathsf B$. The condition $\trc\mathsf B=0$ removes an independent isotropic deformation from the tensorial correction. The scalar isotropic amplitude remains that of the critical AFNS operator.

\section{What the model predicts}

The proposal makes several sharp predictions.

First, the critical limit must be isotropic:
\begin{equation}
h\to\frac13
\quad\Rightarrow\quad
\beta_D\to0
\quad\Rightarrow\quad
\mathsf M\to\II.
\end{equation}
Second, anisotropy should enter to first order as
\begin{equation}
P_{\rm IR}-P_c
\simeq
\eta_c|\xi|^{2/3}\beta_D\bb.
\label{eq:predictionlinear}
\end{equation}
Third, the product of the three principal eigenvalues of the dimensionless tensorial deformation must remain unity:
\begin{equation}
\prod_{i=1}^3 e^{\beta_D\lambda_i}=1.
\label{eq:eigenproduct}
\end{equation}
The model therefore predicts redistribution among directions rather than a new isotropic amplitude.

Fourth, the scale dependence of the tensorial correction is constrained by the same $\beta_D$ that controls the defect scaling. A measurement of $D_\ell$ and $\bb_\ell$ at two or more scales can therefore test the proposed relation without fitting an arbitrary tensor coefficient at each scale.

\section{Falsifiability}

The strongest test of the construction is not whether it can reproduce a selected turbulence statistic after fitting. It is whether the operator relation survives simultaneous tests of its scalar and tensorial predictions.

\begin{enumerate}
\item \textbf{Critical isotropization.} In a regime approaching $h=1/3$, the anisotropic correction inferred from the dissipative response should vanish with $\beta_D$. A finite anisotropic correction at $\beta_D=0$ falsifies the construction.

\item \textbf{Linear critical response.} Close to the threshold, the leading anisotropic response should be proportional to $\beta_D\bb_\ell$. A different leading tensor structure falsifies the minimal closure.

\item \textbf{Unit determinant.} The three principal response factors should satisfy Eq.~\eqref{eq:eigenproduct}. A systematic change in their product would indicate an additional scalar constitutive degree of freedom.

\item \textbf{Two-scale covariance.} Measurements at $\ell$ and $2\ell$ should be compatible with the same constitutive map $e^{\beta_D\bb}$. If independent coefficients are required at each scale, the proposed RG-style construction fails.

\item \textbf{Realizability.} The inferred $\bb_\ell$ must remain inside the realizable Reynolds-stress domain. Persistent excursions outside it falsify the identification of $\bb$ with the normalized coarse-grained covariance.

\item \textbf{Flux consistency.} The mean inertial transfer must remain compatible with the appropriate third-order flux relation. The $4/5$ law should be treated as a constraint on the mean transfer, not as evidence for the local constitutive form.
\end{enumerate}

These tests can be performed in DNS first, where velocity fields and coarse-grained stresses are available at arbitrary filter scales, and subsequently in laboratory flows where sufficiently resolved velocity measurements permit reconstruction of the relevant structure functions and covariance tensors. No claim is made here that these tests have already been passed.

\section{Relation to singular concentration and physical regularization}

The motivation for this construction is partly methodological. Mathematical solutions or approximations can exhibit increasingly concentrated structures as a control parameter changes. Such concentration is an important diagnostic of nonlinear PDE behavior, but it does not by itself establish that the corresponding limiting mechanism is realized by a physical turbulent fluid.

We use this observation to formulate an inverse question. If physical turbulence does not sustain arbitrary concentration at the smallest scales, what macroscopic signature could express the suppression without inserting an ad hoc cutoff? The present proposal answers in two linked steps: the AFNS trajectory changes the scalar order until the Onsager threshold is reached, and the generalized free energy then opens a nonnegative concentration branch only for $h<1/3$. The normalized coarse-grained stress supplies the tensorial information used to continue the critical operator on that branch.

This is a hypothesis about physical fluids, not a conclusion drawn from any particular recent singularity construction. The appropriate arbiter is experiment. If physical measurements show concentration consistent with the proposed tensorial continuation, the model gains support. If they show a different response, the model must be modified or rejected.

\section{Discussion}

The central result can be summarized by the factorization
\begin{equation}
\boxed{
P_{\rm IR}(x,\xi,t)
=\underbrace{\eta_c|\xi|^{2/3}}_{\text{critical AFNS scale}}
\underbrace{e^{\beta_D(x,t)\bb_\ell(x,t)}}_{\text{tensorial deformation}}.
}
\label{eq:factorization}
\end{equation}
The first factor carries the fractional spectral order inherited from the AFNS correspondence trajectory. The second factor carries directional information extracted from the flow itself. The scalar coefficient $\beta_D$ measures the departure from the Onsager threshold, while $\bb_\ell$ supplies the dimensionless anisotropic shape. The generalized free energy adds a separate scalar branch variable $c$ without changing this operator factorization: $c=0$ above the defect threshold and $c>0$ below it. It is therefore a branch selector for the mosaic, not an additional isotropic dissipation amplitude.

Several aspects of the construction are established mathematics, while others are explicitly hypotheses. The definition and scaling of the coarse-grained stress are standard. The existence of a critical $h=1/3$ threshold for inertial energy transfer is part of the Onsager--Duchon--Robert framework. Objective tensor representation and realizability constraints are classical results of turbulence modeling. The separation between the operator beta function $\beta_\delta$ and defect beta function $\beta_D$ follows directly from their definitions. The variational extension uses the standard Landau free-energy logic for an order parameter and the Kuhn--Tucker optimality conditions for a nonnegative boundary variable \cite{Landau1937,GinzburgLandau1950,KuhnTucker1951}; what is new and unproved is the identification of $c$ with a concentration branch and its coupling to $\beta_D$.

By contrast, the choice
\begin{equation}
\mathsf M=e^{\beta_D\bb}
\end{equation}
is a constitutive hypothesis. The exponential is selected because it is the minimal positive-definite parametrization of a symmetric traceless generator, because it reduces to the identity at the critical manifold, and because it does not introduce a new isotropic amplitude. These arguments establish consistency, not uniqueness.

There is also an important limitation concerning locality. The proposed operator is pseudodifferential, and $\bb_\ell$ is scale dependent. A rigorous theory for variable $\bb_\ell(x,t)$ must control the symbol class, commutators, pressure projection, and nonlinear energy estimates. Those questions are separate from the algebraic construction presented here and are not resolved by the present paper.

Finally, intermittency has not been used as a fitting device. Empirical multifractal laws such as She--L\'ev\^eque-type parameterizations may eventually provide stringent comparisons, but importing them into $G_1$ or $G_2$ at the construction stage would obscure whether the operator is genuinely predictive. We therefore leave intermittency statistics as tests of the model rather than as constitutive input.

\section{Conclusions}

We have proposed a compact tensorial continuation of a fractional macroscopic dissipation model. The construction starts from the AFNS trajectory in which the fractional order evolves according to the local Reynolds number and approaches $s=1/3$ at a distinguished scale $\Rey_c$. At that point the isotropic operator is
\begin{equation}
\mathcal L_c=\eta_c(-\Delta)^{1/3}.
\end{equation}

The continuation into the Onsager-critical and subcritical regime is based on two quantities already generated by the coarse-grained flow: the normalized tensor $\bb_\ell$ and the defect beta function $\beta_D$. In addition, the generalized free energy introduces one nonnegative concentration order parameter $c$ and selects the branch $c_*=0$ for $\beta_D\ge0$ and $c_*>0$ for $\beta_D<0$. The resulting operator is
\begin{equation}
\boxed{
\mathcal L_{\rm IR}
=\eta_c\Op^W\!\left[|\xi|^{2/3}e^{\beta_D\bb_\ell}\right].
}
\end{equation}
It recovers the critical AFNS operator exactly when $\beta_D=0$, remains positive at the level of its principal symbol, and separates scalar spectral evolution from tensorial redistribution.

The construction should not be interpreted as a proof that physical turbulence obeys this constitutive law. Its status is that of a constrained, minimal hypothesis. Its scientific value depends on whether it survives tests that do not enter its derivation: DNS, laboratory measurements, two-scale consistency, realizability, and independent energy-flux diagnostics.

The most useful outcome may therefore be the falsifiable statement itself. If physical turbulence exhibits a tensorial response of the form predicted here, the result would support the idea that the Onsager threshold can be represented as a continuation of a macroscopic fractional operator rather than as a separate closure problem. If it does not, the construction identifies precisely where the hypothesis fails. In either case, the experiment remains the final judge.

\appendix
\section{Scaling of the coarse-grained stress and defect}

This appendix derives the scaling exponents quoted in Eqs.~\eqref{eq:theta-regimes} and \eqref{eq:betad}. For a filtered velocity field, the stress in Eq.~\eqref{eq:tau} can be written in terms of velocity increments as a weighted second-order moment. If $|\delta_\ell u|\sim\ell^h$, then dimensional homogeneity gives
\begin{equation}
\tau_\ell\sim\ell^{2h}.
\end{equation}
Multiplication by $\ell^{-2/3}$ gives Eq.~\eqref{eq:theta} and Eq.~\eqref{eq:theta-regimes}.

The coarse-grained energy balance contains the flux
\begin{equation}
\Pi_\ell=-\nabla\bar\uu_\ell:\tau_\ell,
\end{equation}
up to the conventional sign associated with the definition of transfer. Since $\nabla\bar u_\ell\sim\ell^{h-1}$, one obtains
\begin{equation}
\Pi_\ell\sim\ell^{3h-1}.
\end{equation}
The Duchon--Robert defect has the same dimensional scaling. Thus Eq.~\eqref{eq:betad} follows without requiring a pointwise version of the $4/5$ law.

\section{Representation-theoretic reduction}

This appendix derives Eq.~\eqref{eq:minbasis} in detail. Let $\bb$ denote any symmetric traceless $3\times3$ tensor, such as the normalized anisotropy tensor $\bb_\ell$ of Eq.~\eqref{eq:bdef}. By the Cayley--Hamilton theorem, $\bb$ satisfies its own characteristic polynomial,
\begin{equation}
\bb^3-\frac12\trc(\bb^2)\bb-\frac13\trc(\bb^3)\II=0.
\end{equation}
Consequently every polynomial in $\bb$ can be reduced to a linear combination of $\II$, $\bb$, and $\bb^2$. An isotropic tensor-valued function of $\bb$ therefore has the general polynomial form
\begin{equation}
F(\bb)=a_0 \II+a_1 \bb+a_2\bb^2,
\end{equation}
where the scalar coefficients $a_0$, $a_1$, $a_2$ depend only on the independent invariants $\trc(\bb^2)$ and $\trc(\bb^3)$. Taking the traceless part removes the independent isotropic contribution $a_0\II$ together with the isotropic part of $a_2\bb^2$, leaving precisely the basis of Eq.~\eqref{eq:minbasis}.

This is the reduced version of the more general representation theory used in Reynolds-stress closures, in which both strain and rotation tensors are retained and a larger finite tensor basis appears \cite{Pope1975}. The present reduction is not intended to reproduce the full velocity-gradient dependence of conventional closures; it is deliberately restricted to the information carried by the normalized coarse-grained stress.

\section{Positivity and the exponential parametrization}

This appendix establishes the properties of $\mathsf M_\ell=\exp(\beta_D\bb_\ell)$ used in Eqs.~\eqref{eq:M}--\eqref{eq:detM}. Since $\bb=\bb^T$ is real symmetric, the spectral theorem gives an orthogonal diagonalization
\begin{equation}
\bb=O\Lambda O^T,
\qquad
\Lambda=\operatorname{diag}(\lambda_1,\lambda_2,\lambda_3),
\end{equation}
with $O^TO=\II$ and real eigenvalues $\lambda_1,\lambda_2,\lambda_3$. Because the matrix exponential commutes with orthogonal conjugation,
\begin{equation}
e^{\beta_D \bb}=O\operatorname{diag}(e^{\beta_D\lambda_1},e^{\beta_D\lambda_2},e^{\beta_D\lambda_3})O^T.
\end{equation}
Each factor $e^{\beta_D\lambda_i}$ is strictly positive for any real $\beta_D$ and any real $\lambda_i$, so $e^{\beta_D\bb}$ is symmetric positive definite regardless of the sign or magnitude of $\beta_D$. Furthermore, since $\trc\bb=\lambda_1+\lambda_2+\lambda_3=0$,
\begin{equation}
\det(e^{\beta_D \bb})=e^{\beta_D\trc \bb}=e^{\beta_D(\lambda_1+\lambda_2+\lambda_3)}=1.
\end{equation}
Thus the exponential parametrization is a convenient closed-form way of enforcing positivity for arbitrarily large anisotropy while preserving the volume of the tensorial deformation in eigenvalue space, exactly as stated in Eqs.~\eqref{eq:detM} and \eqref{eq:eigenproduct}.

The result does not by itself prove a global energy inequality for the full variable-coefficient pseudodifferential equation. Such a proof would require estimates on the Weyl quantization and its lower-order terms. The statement established here is positivity of the principal symbol.

\section{Two-scale consistency}

Let $\mathcal R_a$ denote a scale transformation $\ell\mapsto a\ell$. The constitutive prescription is
\begin{equation}
\mathcal R_a:\quad
(\delta_\ell,\beta_{D,\ell},\bb_\ell)
\mapsto
(\delta_{a\ell},\beta_{D,a\ell},\bb_{a\ell}),
\end{equation}
while the same function
\begin{equation}
\mathcal F(\beta_D,\bb)=e^{\beta_D \bb}
\end{equation}
constructs the tensorial response at every scale. This is a constitutive scale covariance condition. It is weaker than a full Wilsonian RG because no path-integral measure, mode elimination, or complete theory-space transformation is specified. The terminology ``RG-style'' is therefore more precise than claiming a complete Wilsonian derivation.

\section{Connection with the mean $4/5$ law}

For homogeneous isotropic turbulence in the inertial range,
\begin{equation}
\left\langle(\delta u_L)^3\right\rangle=-\frac45\varepsilon\ell.
\end{equation}
This implies a scale-independent mean energy flux. It does not imply that the local Duchon--Robert defect is constant. The present construction uses the $4/5$ law only as an external consistency condition on the mean transfer, while $\beta_D$ describes local scaling. This distinction is essential if the model is to remain compatible with intermittency.

\section{Limits of the construction}

Six limitations should be kept explicit. First, the AFNS interpolation in Eq.~\eqref{eq:sdelta} is a model choice rather than a microscopic derivation. Second, the identity $h-1/3=\kappa-2\delta$ is an imposed constitutive relation of the AFNS framework and requires independent validation. Third, the identification of $c$ as a concentration order parameter and the linear coupling $g\beta_Dc$ are phenomenological hypotheses, even though their variational structure is standard. Fourth, the parameters $a$ and $g$ have not been independently determined from data. Fifth, the exponential tensorial closure is a minimal hypothesis rather than a theorem. Sixth, variable-coefficient fractional operators require analytical control beyond the principal-symbol calculation presented here.

These limitations are not technical footnotes: they define the empirical content of the proposal. The model is useful only if its predictions can be separated from its assumptions in numerical or laboratory tests.

\section{The constitutive bridge between fractional order, Reynolds scaling, and velocity regularity}

This appendix makes explicit a point that is central to the construction but easy to obscure in a compact sequence of equations: the defect-scaling relations and the AFNS operator flow are distinct, and the relation connecting them is a constitutive assumption of the AFNS framework rather than an algebraic consequence of the preceding definitions.

We begin with the local increment scaling
\begin{equation}
|\delta_\ell u|\sim \ell^h,
\label{eq:app_increment}
\end{equation}
where $h$ is the effective velocity-regularity exponent over the scale window under consideration. The Duchon--Robert defect is cubic in velocity increments and contains one inverse power of the coarse-graining length. At the level of scaling, therefore,
\begin{equation}
D_\ell\sim\frac{|\delta_\ell u|^3}{\ell}
\sim\ell^{3h-1}.
\label{eq:app_D}
\end{equation}
This is the origin of the Onsager threshold: the exponent vanishes precisely at $h=1/3$. For $h>1/3$ the scaling forces the defect to decay toward small scales; for $h=1/3$ it permits a scale-independent defect; and for $h<1/3$ the scaling no longer forces the defect to vanish. These statements are scaling statements and do not assert that every field below the threshold generates a nonzero positive defect.

It is therefore natural to define the defect beta function by
\begin{equation}
\beta_D\equiv\frac{d\ln|D_\ell|}{d\ln\ell}=3h-1.
\label{eq:app_betaD}
\end{equation}
This beta function describes the scale dependence of the inertial defect. It is not the beta function of the AFNS fractional order.

The second quantity is the local Reynolds-number scaling exponent,
\begin{equation}
\kappa\equiv\frac{d\ln\Rey_\ell}{d\ln\ell}.
\label{eq:app_kappa}
\end{equation}
For example, the idealized K41 estimate $u_\ell\sim\ell^{1/3}$ gives $\Rey_\ell\sim\ell^{4/3}$ and hence $\kappa=4/3$. More generally, $\kappa$ is allowed to vary with scale.

The AFNS order is written as
\begin{equation}
s=\frac13+\delta,
\label{eq:app_sdelta}
\end{equation}
so that $2\delta$ measures the displacement of the Fourier order $2s$ from its critical value $2/3$. The AFNS interpolation used in the main text gives
\begin{equation}
\delta(\Rey_\ell)=\frac{2/3}{1+(\Rey_\ell/\Rey_c)^\gamma}.
\label{eq:app_deltaRe}
\end{equation}
Differentiating Eq.~\eqref{eq:app_deltaRe} with respect to $\ln\Rey_\ell$ yields
\begin{equation}
\frac{d\delta}{d\ln\Rey_\ell}
=-\gamma\delta\left(1-\frac32\delta\right).
\label{eq:app_deltare}
\end{equation}
Consequently, by the chain rule,
\begin{equation}
\beta_\delta\equiv\frac{d\delta}{d\ln\ell}
=-\gamma\kappa\delta\left(1-\frac32\delta\right),
\label{eq:app_betadelta}
\end{equation}
which reduces to the K41 expression quoted in the main text when $\kappa=4/3$.

The crucial bridge is the relation
\begin{equation}
\boxed{
h-\frac13=\kappa-2\delta.
}
\label{eq:app_constitutive}
\end{equation}
Equation~\eqref{eq:app_constitutive} is the \emph{constitutive scaling relation of the AFNS framework}. It does not follow algebraically from Eqs.~\eqref{eq:app_D}--\eqref{eq:app_betadelta}, nor is it a consequence of the definition of the Duchon--Robert defect. It is the additional structural hypothesis that connects the operator sector, represented by $s$ or $\delta$, to the geometric regularity sector, represented by $h$, through the scale dependence of the local Reynolds number.

Using $\delta=s-1/3$, Eq.~\eqref{eq:app_constitutive} can equivalently be written as
\begin{equation}
\boxed{
h=\kappa+1-2s.
}
\label{eq:app_constitutive_s}
\end{equation}
This form makes the content of the constitutive assumption particularly transparent. The quantity $\kappa$ measures how the local Reynolds number changes along the scale trajectory, while $2s$ is the differential order of the fractional operator in Fourier space. The relation therefore postulates that the departure of the velocity regularity from the Onsager value is controlled by the difference between these two scale exponents.

Combining Eq.~\eqref{eq:app_constitutive} with Eq.~\eqref{eq:app_betaD} gives, without any further assumption,
\begin{equation}
\boxed{
\beta_D=3(\kappa-2\delta).
}
\label{eq:app_betaDbridge}
\end{equation}
Thus the defect is scale invariant precisely on the manifold
\begin{equation}
\beta_D=0
\quad\Longleftrightarrow\quad
h=\frac13
\quad\Longleftrightarrow\quad
\kappa=2\delta.
\label{eq:app_critical}
\end{equation}
This condition should not be confused with the operator endpoint $s=1/3$, which is equivalent to $\delta=0$. In particular, $s=1/3$ alone does not imply $h=1/3$; Eq.~\eqref{eq:app_constitutive} shows that $h=1/3$ at $\delta=0$ additionally requires $\kappa=0$. The coincidence of the AFNS operator endpoint with the Onsager regularity threshold is therefore a property of the trajectory in the joint $(\delta,\kappa)$ space, not an identity between the two thresholds.

The construction consequently contains two distinct scale flows. The operator flow is
\begin{equation}
\beta_\delta=\frac{d\delta}{d\ln\ell},
\end{equation}
whereas the defect flow is
\begin{equation}
\beta_D=\frac{d\ln|D_\ell|}{d\ln\ell}.
\end{equation}
They should not be identified. The constitutive relation in Eq.~\eqref{eq:app_constitutive} is the bridge between them. This distinction is essential for the interpretation of the generalized operator: the fractional order describes the trajectory of the dissipative operator, while $\beta_D$ measures the scaling departure of inertial energy transfer from the Onsager threshold.

The scientific status of Eq.~\eqref{eq:app_constitutive} is therefore explicit and falsifiable. Given independent estimates of $h$, $\kappa$, and $s$ over the same scale window, the AFNS relation predicts
\begin{equation}
h-\frac13\stackrel{?}{=}\kappa-2\left(s-\frac13\right).
\label{eq:app_test}
\end{equation}
Likewise, an independent measurement of the scale dependence of the inertial defect provides the test
\begin{equation}
\beta_D\stackrel{?}{=}3\left(\kappa-2\delta\right).
\label{eq:app_test_beta}
\end{equation}
Failure of these relations would falsify the constitutive bridge while leaving the standard Duchon--Robert scaling and the definition of the AFNS operator intact. Conversely, agreement would provide evidence for the specific AFNS mechanism connecting fractional operator order to velocity regularity.

The role of this appendix is consequently not to add another closure assumption. It is to isolate the one assumption in the scaling chain that carries genuinely new physical content. Equations~\eqref{eq:app_D} and \eqref{eq:app_betaD} express the defect scaling; Eq.~\eqref{eq:app_betadelta} expresses the AFNS order flow; Eq.~\eqref{eq:app_constitutive} is the constitutive bridge between the two sectors. Making this distinction explicit is necessary for a precise interpretation and for a meaningful falsifiability test of the proposed continuation.

\section{A single equation with a regime-selective tensorial continuation}

The preceding construction reaches the critical operator
\begin{equation}
\mathcal L_c=\eta_c(-\Delta)^{1/3},
\qquad \eta_c=\frac{1}{\Rey_c},
\end{equation}
from the scalar AFNS trajectory. The question addressed here is whether the tensorial continuation can be incorporated without introducing a second evolution equation. We formulate the continuation as a regime-selective component of one macroscopic operator equation. The construction is designed so that the scalar AFNS operator governs the ordinary branch, the tensorial operator governs the concentrated branch, and both coincide at the critical interface.

For the scalar AFNS branch, the spectral calculus of the positive operator $-\Delta$ gives
\begin{equation}
(-\Delta)^s=\exp\!\left[s\ln(-\Delta)\right].
\label{eq:logAFNS}
\end{equation}
Thus, with $s=1/3+\delta$,
\begin{equation}
\mathcal L_{\rm AFNS}
=\eta(\rho)\exp\!\left[\frac13\ln(-\Delta)+\delta(\rho)\ln(-\Delta)\right],
\qquad
\rho=\frac{\Rey_\ell}{\Rey_c}.
\label{eq:logAFNSsplit}
\end{equation}
The two terms in the exponent commute because both are functions of the same self-adjoint operator $-\Delta$. Equation~\eqref{eq:logAFNSsplit} is therefore exactly equivalent to the fractional AFNS operator; it is not an additional approximation.

At the endpoint $\delta\to0$, the scalar operator reaches $\mathcal L_c$. Beyond that interface, the proposed tensorial continuation is represented by the Weyl-quantized operator
\begin{equation}
\mathcal L_{\rm IR}
=\eta_c\operatorname{Op}^{W}
\!\left[|\xi|^{2/3}\exp(\beta_D\bb_\ell)\right].
\label{eq:LIRappendix}
\end{equation}
The use of Weyl quantization is standard for spatially dependent pseudodifferential symbols and fixes the operator ordering through the midpoint prescription \cite{Hormander1979,Folland2016}. It is important that Eq.~\eqref{eq:LIRappendix} is an operator defined by its symbol; it is not obtained by naively multiplying $(-\Delta)^{1/3}$ by a position-dependent matrix, since in general
\begin{equation}
[(-\Delta)^{1/3},\bb_\ell(x,t)]\ne0.
\end{equation}

The two sectors must not be activated by an independent phenomenological switch. The concentration variable introduced by the generalized free-energy construction provides the appropriate state variable. Let $c\ge0$ denote the dimensionless concentration amplitude and consider the local potential
\begin{equation}
\mathcal F(f,c;\rho,\kappa)
=F_0(f;\rho)+\frac{a}{2}c^2+g\,\beta_D(f,\kappa)c,
\qquad a>0,
\label{eq:Fswitch}
\end{equation}
where $F_0$ is the original AFNS regularity free energy and
\begin{equation}
\beta_D=3(\kappa-2\delta)=3\kappa-4f,
\qquad
\delta=\frac23 f.
\label{eq:betafswitch}
\end{equation}
Minimization with respect to $c$ under the constraint $c\ge0$ gives the Kuhn--Tucker solution
\begin{equation}
c_* = \frac{g}{a}\,[-\beta_D]_+,
\qquad [x]_+=\max(x,0),
\label{eq:cstarH}
\end{equation}
for $g>0$. Hence $c_*=0$ on the non-concentrated branch $\beta_D\ge0$, while $c_*>0$ precisely when $\beta_D<0$, i.e. when
\begin{equation}
h<\frac13.
\label{eq:concentrationbranch}
\end{equation}
The concentration variable therefore does not constitute a new independent physical criterion: its activation is determined by the same constitutive bridge that defines the Onsager-critical manifold.

To turn this order parameter into a bounded operator weight without introducing another dimensional scale, define the monotone activation function
\begin{equation}
\chi=\frac{c_*}{1+c_*},
\qquad 0\le\chi<1.
\label{eq:chi}
\end{equation}
Thus $\chi=0$ whenever the concentration branch is inactive and $\chi$ increases monotonically with the strength of concentration. The upper bound is not intended to assert a saturated physical concentration; it only guarantees that the operator interpolation remains a convex combination.

The single macroscopic equation is then
\begin{equation}
\boxed{\begin{gathered}
\partial_t\uu+(\uu\cdot\nabla)\uu+\nabla p
+\left[(1-\chi)\mathcal L_{\rm AFNS}+\chi\mathcal L_{\rm IR}\right]\uu=0,\\[3pt]
\nabla\cdot\uu=0.
\end{gathered}}
\label{eq:singlemother}
\end{equation}
Equation~\eqref{eq:singlemother} contains two operator sectors, but only one dynamical equation. The scalar AFNS sector is active when $c_*=0$; the tensorial sector is activated continuously by the concentration order parameter. No second velocity field, auxiliary evolution equation, or independent dissipation coefficient is introduced.

The construction is consistent at the interface. When $\beta_D=0$, one has $c_*=0$ and therefore $\chi=0$. At the same point the tensorial generator vanishes,
\begin{equation}
\beta_D\bb_\ell=0,
\end{equation}
so that
\begin{equation}
\mathcal L_{\rm IR}
=\eta_c\operatorname{Op}^{W}\!\left[|\xi|^{2/3}I\right]
=\eta_c(-\Delta)^{1/3}
=\mathcal L_c.
\label{eq:operatorinterface}
\end{equation}
The scalar AFNS trajectory reaches the same operator at its critical endpoint. Consequently the two descriptions are joined through the common operator $\mathcal L_c$, rather than through an arbitrary matching condition.

The limiting regimes are now immediate. On the ordinary branch, $c_*=0$ and $\chi=0$, so
\begin{equation}
\mathcal L=\mathcal L_{\rm AFNS}.
\end{equation}
As $\Rey_\ell/\Rey_c\to0$, $s\to1$ and the AFNS operator reduces to the Navier--Stokes Laplacian with the corresponding viscous amplitude. At the critical endpoint, $s\to1/3$ and $\eta\to\eta_c$, giving $\mathcal L_c$. On the concentrated branch, $c_*>0$ and $\chi>0$, so the tensorial continuation becomes dynamically active and progressively replaces the scalar operator as concentration strengthens.

The essential structural feature is therefore a change in the character of the constitutive generator along a single Reynolds-scale journey. Before the critical interface, the generator is scalar,
\begin{equation}
\mathcal G_{\rm AFNS}
=\frac13\ln(-\Delta)I+\delta\ln(-\Delta)I,
\label{eq:Gscalar}
\end{equation}
where the deformation is purely spectral. On the concentrated side, the continuation is generated by
\begin{equation}
\mathcal G_{\rm IR}
=\frac13\ln(-\Delta)I+\beta_D\bb_\ell,
\label{eq:Gtensor}
\end{equation}
with the second term now tensorial and spatially dependent. The journey therefore changes from a scalar deformation of the fractional order to a tensorial deformation of the critical operator. Weyl quantization is the operator-level realization of this second stage; it does not constitute a separate physical closure.

This formulation also clarifies why the two operators should not simply be added with fixed coefficients. A fixed sum would retain both sectors in every regime and would generally double-count the critical response. Equation~\eqref{eq:singlemother} instead uses the concentration state to determine which sector contributes. The equality in Eq.~\eqref{eq:operatorinterface} guarantees that the interpolation has no artificial jump at the critical interface.

The status of the construction is deliberately limited. The AFNS interpolation, the constitutive relation $h-1/3=\kappa-2\delta$, and the exponential tensorial response remain hypotheses requiring independent tests. The new single-equation formulation adds no claim beyond their combination: it provides a mathematically explicit way of embedding the scalar AFNS branch and its tensorial continuation in one macroscopic evolution equation, with a common critical operator and a state-dependent activation determined by the same concentration criterion. In particular, the formulation is falsifiable if the observed onset of concentration fails to correlate with the predicted $\beta_D<0$ branch, or if the measured tensorial response is incompatible with the proposed Weyl symbol.

\end{document}